\documentclass[aps,pra,epsfigure,twocolumn,longbibliography,superscriptaddress]{revtex4-1}

\usepackage[T1]{fontenc}
\usepackage[utf8]{inputenc}
\usepackage{lmodern}
\usepackage[english]{babel}

\usepackage{graphicx}
\usepackage{epstopdf}
\usepackage{epsfig,times}
\usepackage{dcolumn}
\usepackage{bm,natbib,mathrsfs}
\usepackage{color}
\usepackage{amssymb}
\usepackage{amsfonts}
\usepackage{amsmath}
\usepackage{amstext}
\usepackage{latexsym}
\usepackage{physics}
\usepackage[colorlinks=true,linkcolor=blue,urlcolor=blue,citecolor=blue,pdfusetitle]{hyperref}
\usepackage{hyperref}
\usepackage[dvipsnames]{xcolor}

\usepackage{dcolumn}    
\usepackage{bm} 
\usepackage{graphicx}
\usepackage{amsmath}    
\usepackage{latexsym}
\usepackage{amsfonts}   
\usepackage{amssymb}
\usepackage{array}      
\usepackage{epsfig}
\usepackage{txfonts}
\usepackage{color}
\usepackage[colorlinks=true,linkcolor=blue,urlcolor=blue,citecolor=blue,pdfusetitle]{hyperref}
\usepackage{hyperref}
\usepackage[dvipsnames]{xcolor}

\newcommand{\beq}{\begin{equation}}
\newcommand{\eeq}{\end{equation}}
\newcommand{\bea}{\begin{eqnarray}}
\newcommand{\eea}{\end{eqnarray}}

\newcommand{\om}{\omega} 
\newcommand{\Om}{\Omega}

\newcommand{\dw}{\downarrow}
\newcommand{\QD}{Quantum Darwinism}
\newcommand{\up}{\uparrow}
\DeclareMathOperator*{\Motimes}{\text{\raisebox{0.25ex}{\scalebox{0.8}{$\bigotimes$}}}}
\begin{document}
\author{Salvatore Lorenzo}
\affiliation{Dipartimento di Fisica e Chimica - Emilio Segr\'e, Universit\`a  degli Studi di Palermo, via Archirafi 36, I-90123 Palermo, Italy}
\author{Mauro Paternostro}
\affiliation{Centre for Theoretical Atomic, Molecular, and Optical Physics, School of Mathematics and Physics, Queen's University, Belfast BT7 1NN, United Kingdom}
\author{G. Massimo Palma}
\affiliation{ Dipartimento di Fisica e Chimica - Emilio Segr\`e ,Universit\`a degli Studi di Palermo,
via Archirafi 36, I-90123 Palermo, Italy}
\affiliation{NEST, Istituto Nanoscienze-CNR, Piazza S. Silvestro 12, 56127 Pisa, Italy}

\begin{abstract}
We combine the collisional picture for open system dynamics and the control of the rate of decoherence provided by the quantum (anti-)Zeno effect to illustrate the temporal unfolding of the redundant encoding of information into a multipartite environment that is at the basis of Quantum Darwinism, and to control it. The rate at which such encoding occurs can be enhanced or suppressed by tuning the dynamical conditions of system-environment interaction in a suitable and remarkably simple manner.  This would help the design of a new generation of quantum experiments addressing the elusive phenomenology of \QD~ and thus its characterization.
\end{abstract}

\title{ (Anti-)Zeno-based dynamical control of the unfolding of quantum Darwinism}
\maketitle

Quantum Darwinism (QD) is an interesting theoretical framework that strives at explaining the emergence of objective reality out of quantum superpositions, a most fundamental questions in modern quantum theory, in terms of the proliferation or redundant records of the quantum state of a quantum system in the environment \cite{zurek_NP2009}. 

The basic idea is that, due to their joint interaction, an environment gets entangled with the system, thus acquiring information about its state \cite{zurek_RMP2003,blume-kohout_FP2005}. The set of states which get entangled with the environment, the so called pointer states, are the eigenstates of the observable of the system that enters the coupling with the environment~\cite{zurek_PRD1981,zurek_PRD1982}. Pointer states are left undisturbed by the mutual interaction with the environment, but their coherent superposition gets entangled with it. Any external observer who can access the environment can thus acquire information about the system~\cite{korbicz_PRL2014,horodecki_PRA2015}. 

A second important assumption of the \QD~model is that the environment is not a single block but rather a collection of independent units, each of which endowed with accessible information about the system state, which has been redundantly encoded in the environment through the entangling process mentioned above~\cite{zwolak_PRA2010,zwolak_PRA2017,zwolak_SR2016}. Different observers having access to separate environmental fragments will have access to the same shared information about the system, which will thus become an element of objective reality, and thus an inherently classical quantity~\cite{horodecki_PRA2015,le_PRA2018}. More formally, the assumption is that an initial coherent superposition of  pointer states of the system $|\Psi\rangle_S = \sum_{k} \psi_k |\pi_k\rangle_S$ evolves into a joint system-environment state with a branching structure  
\begin{equation}
\label{uno}
|\Psi_{SE}\rangle = \sum_{k} \psi_k |\pi_k\rangle_S\bigotimes^n_{j=1}|\eta_k\rangle_j,
\end{equation}
where $n$ is the number of elements of the environment. Eq.~\eqref{uno} shows that the information about the system pointer state $|\pi_k\rangle_S$ is imprinted into multiple copies of environmental states $|\eta_k\rangle$, thus becoming accessible to individual, distinct observers, that access separate fragments of the environment. 

The emergence of \QD~has been extensively studied in recent literature, both theoretically and experimentally~\cite{ciampini_PRA2018,burke_PRL2010,brunner_PRL2008,chen_AQ2018,unden_AQ2018,Quanta}, and its links with other phenomena characterising quantum and open quantum system, such as non-Markovianity~\cite{lampo_PRA2017,pleasance_PRA2017,giorgi_PRA2015,galve_SR2016,Milazzo2019}, non-contextuality inequalities~\cite{Baldijao2018}, and spectrum broadcasting structures~\cite{korbicz_PRL2014} have been addressed and studied. Remarkably, while \QD~implies spectrum broadcasting, the reverse is not the case, which manifests the only partial understanding that we currently have of the interplay between Quantum Darwinism, quantum correlations, and broadcasting structures~\cite{horodecki_PRA2015, zwolak_PRL2009, zwolak_PRA2010,le_PRA2018,le_PRL2019,lampo_PRA2017,Mironowicz_PRA2018,Mironowicz_PRL2017}.

Notwithstanding such attention, the phenomenology of \QD~ an its role in the occurrence of the quantum-to-classical transition are yet to be understood fully~\cite{Campbell2019}, particularly in relation to the mechanism of its temporal unfolding. In this work we address this issue by adopting a collision-model approach to open quantum system dynamics~\cite{ziman_PRA2002,ziman_PRA2005,ziman_OSID2005,ziman_QDI2010} in which the interaction with the environment, consisting of an infinite number of identical elements dubbed ancillas, takes place through a sequence of rapid interactions (collisions) between system and ancillas. This is very close to real physical situations, like, e.g. the learning about the state of a system by observing the photons that are scattered by it \cite{ollivier_PRL2004} and has been recently successfully used to dig into the Darwinistic phenomenology. We show that not only this approach provides a transparent picture of the process of successive redundant encoding of information that ultimately results in the emergence of objective reality but also it allows for the design of simple control strategies for the harnessing of \QD~itself. We make use of the dynamical version of the quantum Zeno effect~\cite{misra_JoMP1977}, which does not rely on fast measurements but on cleverly arranged dynamical conditions~\cite{peres_AJoP1980}. Through this, we show that the rate at which information is spread across an environment  -- consisting of many units that interact with a system  by collisions -- can be reduced or enhanced, an thus ultimately controlled. In turn, we believe that such (remarkably simple) control strategies can be fruitfully exploited for further experimental investigations along the lines of Refs.~\cite{ciampini_PRA2018,burke_PRL2010,brunner_PRL2008,chen_AQ2018,unden_AQ2018,Quanta}, and thus used to critically address the role of \QD~in the transition of quantumm systems to classicality.

The remainder of this manuscript is structured as follows. In Sec.~\ref{unfolding} we introduce our collision model of pure decoherence and derive the unfolding of \QD. Sec.~\ref{controlling} is dedicated to the description of our approach to the control of such unfolding: we show that the judicious arrangement of a Quantum Zeno mechanism is effective in inhibiting the spreading of redundant information from the system to fractions of the environment (cf. Subsec.~\ref{Zeno}), while an anti-Zeno-like effect accelerate such effect towards a faster onset of \QD~[cf. Subsec.~\ref{AntiZeno}]. Finally, in Sec.~\ref{conc} we draw the conclusions that can be gathered from our work and briefly address the avenues that they open.

\section{Unfolding of Darwinism in a memoryless collision model of decoherence}
\label{unfolding}
In order to illustrate how \QD~unfolds in time we address decoherence resulting from a simple memoryless collision model. While providing an intuitive mechanism for \QD, the collisional picture is rich enough to allow for the grasping of the subtleties associated to the emergence of objective reality.  

\begin{figure}[!t]
	\begin{center}
		\includegraphics[width=0.275\columnwidth]{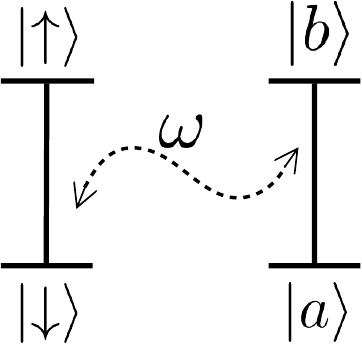}\qquad\qquad
		\includegraphics[width=0.45\columnwidth]{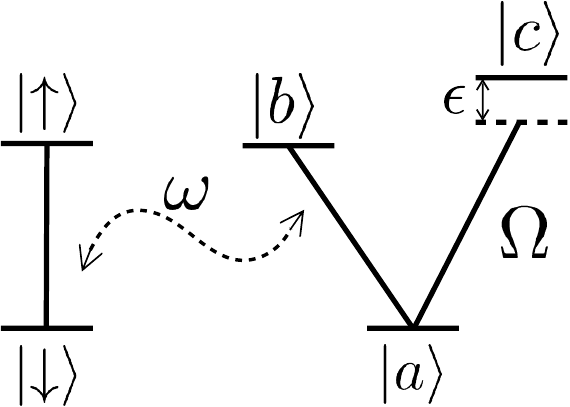}  
		\qquad\qquad\qquad\qquad\textbf{(a)} \qquad\qquad\qquad\qquad\qquad  \qquad \textbf{(b)}    \qquad\qquad
		\caption{{(Color online) Schematic representations of the interaction models for the temporal unfolding of \QD~considered in this work. Panel {\bf(a)} illustrates the configuration of system-environment interactions that give rise to redundant encoding of information about the system onto the state of the environmental elements via repeated collisions. Each involves subenvironments embodied by two-level-systems interacting with ${\cal S}$ according to the Hamiltonian in Eq.~\eqref{H1}. Panel {\bf(b)} is for the depletion (enhancement) of the rate of redundant encoding achieved through a (anti-)Zeno-like effect. We induce this by choosing three-level-systems as ancillas. While the $\ket{a}\leftrightarrow\ket{c}$ is driven with a (dimensionless) Rabi frequency $\Omega$, the $\ket{a}\leftrightarrow\ket{b}$ one is directly affected by the collision with the system. Different regimes of redundant encoding are achieved by adjusting the detuning $\epsilon$.}}    
		\label{fig_0}
	\end{center}
\end{figure}

As in a standard collisional model of open system dynamics, the environment consists of a large ensemble of subenvironments, dubbed ancillas. The irreversible reduced dynamics of the system is described in terms of a sequence of brief interactions (collisions) between the system and the ancillas. In the memoryless scenario the number of ancillas in the environment is infinite and the system never collides twice with same ancilla~\cite{ciccarello_PRA2013,lorenzo_PRA2017}. If we assume $n\gg1$ and the ancillas all prepared in the same state ${\hat \eta}$, after $\ell$ collisions the joint system-environment state will be ${\hat \rho}_{\mathcal{SR}} = {\hat U}_{\{\ell\}}({\hat\rho_0}\otimes{\hat \eta}^{\otimes n}){\hat U}^\dag_{\{\ell\}}$ where $\hat{U}_{\{\ell\}}={\hat U}_{\ell}{\hat U}_{\ell-1}\cdots {\hat U}_{1}$ and ${\hat U}_k$ describes the collision between the $k^{th}$ ancilla and the system. To be specific, let us now consider a two-level system $\mathcal{S}$ and an environment $\mathcal{R}$, in turn composed by a large collection of two-level systems $\mathcal{R}_n$. The $\mathcal{S}$-$\mathcal{R}$  collisions are generated by the following interaction Hamiltonian [cf. Fig.~\ref{fig_0} {\bf (a)}]
\bea
\hat H_n=\hbar\omega {\hat\sigma}_\mathcal{S}^z\otimes{\hat\sigma}^x_{\mathcal{R}_n},
\label{H1}
\eea
where ${\hat\sigma}^z_\mathcal{S}=\ket{\up}\bra{\up}_{\mathcal{S}}-\ket{\dw}\bra{\dw}_{\mathcal{S}}$ and ${\hat\sigma}^x_{\mathcal{R}_j}=\ket{a}\bra{b}_{\mathcal{R}_j}+\ket{b}\bra{a}_{\mathcal{R}_j}$ are the $z$- and $x$-Pauli matrices for $\mathcal{S}$ and $\mathcal{R}_j$, respectively.
In order to fix the ideas and with no loss of generality we ignore at this stage any free system or ancilla dynamics.
The collisions are thus described by a rotation operator on the ancilla conditioned on the state of the system
\bea
{\hat U}=\exp\{-i\omega {\hat\sigma}_\mathcal{S}^z\otimes{\hat\sigma}^x_{\mathcal{R}_n}\tau\}
\eea
where $\tau$ stands for the collision time, i.e. the duration of each collision.
Such dynamics will induce entanglement between the system and an increasing numbers of ancillas as the collisions take place. Starting from $\Psi_0=(\alpha\ket{\dw}+\beta\ket{\up})\Motimes_{j=1}^{n}\ket{a}$ we can write the $\mathcal{S}$-$\mathcal{R}$ joint state after $\ell$ collisions as
\bea
\ket{\Psi_\ell}=\left(\alpha\ket{\dw}\Motimes_{j=1}^\ell\ket{\varphi^+}_{\mathcal{R}_j}+\beta\ket{\up}\Motimes_{j=1}^\ell\ket{\varphi^-}_{\mathcal{R}_j}\right)\Motimes_{j=\ell{+}1}^n\ket{a}_{\mathcal{R}_j},
\eea
where $\ket{\varphi^\pm}=\cos(\om\tau)\ket{a}\pm i\sin(\om\tau)\ket{b}$.
The corresponding reduced density matrix $\rho_{\mathcal{S}}^\ell$ is
\bea
\rho_{\mathcal{S}}^\ell=\tr_\mathcal{R}\{\ket{\Psi_\ell}\bra{\Psi_\ell}\}=\left(\begin{array}{cc}   |\alpha|^2 & \alpha\beta^*\kappa^\ell\\ \alpha^*\beta\kappa^{*\ell} & |\beta|^2\end{array}\right)
\label{reduced}\eea
with $\kappa=|\braket{\varphi^-}{\varphi^+}|=|\cos(2\om\tau)|$.
The above expressions show clearly that the states $\ket{\dw}_{\!\mathcal{S}}$ and $\ket{\up}_{\!\mathcal{S}}$ are pointer states and the ancillas, collision after collision, gradually acquire information about the system. 

{Tipically many collisions, i.e. many subenvironments, are necessary to distinguish pointer states. The number of times that this information can be independently extracted is taken as measure of objectivity. The mutual information between the system and a subset $m$ of ancillas after $\ell$ collisions
\bea
\mathcal{I}(\mathcal{S},\mathcal{F}^\ell_m)=
S(\rho_{\mathcal{S}}^{\ell})+S(\rho_{\mathcal{F}_m}^\ell)-S(\rho_{\mathcal{SF}_m}^\ell)
\eea
is the information about $\mathcal{S}$ available from $\mathcal{F}_m$. We call ${\cal F}^\ell_{\delta}$ the smallest fraction of the environment such that, for any arbitrarily picked information gap $\delta$, we achieve 
\bea
\mathcal{I}(\mathcal{S},\mathcal{F}^\ell_{\delta})\geq(1-\delta)S(\rho_{\mathcal{S}}^{\ell}).
\label{deficit}\eea
We can thus define the redundancy as 
\bea
R{=}\ell/\mathcal{F}^\ell_\delta,
\eea}
which provides a quantitative estimate of the number of distinct subenvironments that supply classical information about ${\cal S}$, up to the deficit $\delta$.

The off diagonal matrix elements of $\rho^\ell_\mathcal{S}$  decay as $\kappa^{\ell} = e^{-\Gamma\ell}$ where we have defined the decay constant $\Gamma = - \ln\kappa$. After ${\tilde \ell } = \Gamma^{-1}$ collisions, $ \rho_{\mathcal{S}}^{\tilde\ell}$ will be fully decohered. {Further collisions will thus take place with the system that is in a state with no quantum coherence}. This  implies that the size of environmental fragments which have nearly complete knowledge about the state of the system is ${\cal O}({\tilde\ell^{-1}})$ and, after $n$ collisions, we expect a redundancy $R \sim - n\log \kappa$. We will show shortly the correctness of this estimate. 

Before proceeding to the analytic evaluation of the quantum mutual information, let us point out that, in the limit of very short collision time that justifies a limit to continuous time, the above dynamics can be effectively described by the dephasing master equation
\bea
\frac{d\rho}{dt}=\gamma(\hat \sigma^z\rho\hat\sigma^z-\rho)
\label{ME1}\eea
with $\gamma=\om^2\tau$ the decoherence rate.
To evaluate the mutual information between the system and growing environmental fragments, we make extensive use of the fact that the overall system-environment state remains pure after an arbitrary number of collisions.
The von Neumann entropy of $\mathcal{S}$ is easily calculated at each step as $S(\rho_{\mathcal{S}}^\ell)=-\sum_{j=\pm} \lambda_j\log_2\lambda_j$ with $\lambda_\pm=(1\pm\sqrt{1-4(1-|\kappa|^{2\ell})|\alpha|^2|\beta|^2})/2$.
We now need the quantum von Neumann entropies of $\rho_{\mathcal{F}_m}$ and $\rho_{\mathcal{SF}_m}$.
Let us note that, after $\ell>m$ collisions,  the reduced density matrix of the fragment ${\mathcal{F}_m}$ is given by
\bea
\rho_{\mathcal{F}_m}^{\ell>m}=|\alpha|^2 \Motimes_{j=1}^m\ket{\varphi^+}\bra{\varphi^+}_{\mathcal{R}_j}+|\beta|^2 \Motimes_{j=1}^m\ket{\varphi^-}\bra{\varphi^-}_{\mathcal{R}_j}.
\eea
This means that after the first $m$ collisions the reduced density matrix of $\mathcal{F}_m$ does not change further and, recalling that within the first $m$ collisions the joint $\mathcal{S}$-$\mathcal{F}_m$ state is pure, we have that
\bea
S(\rho_{\mathcal{F}_m}^{\ell>m})=S(\rho_{\mathcal{S}}^m).
\eea
With similar considerations, the entropy of $\rho_{\mathcal{SF}_m}$ turns out  to be
$S(\rho_{\mathcal{SF}_m}^{\ell>m})=S(\rho_{\mathcal{S}}^{\ell-m})$.
On other hand, if $\ell<m$ we have that 
$S(\rho_{\mathcal{F}_m}^{\ell<m})=S(\rho_{\mathcal{S}}^{\ell<m})$, and $S(\rho_{\mathcal{SF}_m}^{\ell<m})=0$.
Putting toghether all the above relations, the mutual information between $\mathcal{S}$ and $\mathcal{F}_m$ is
\bea
\mathcal{I}(\mathcal{S},\mathcal{F}^\ell_m)=
\begin{cases}
S(\rho_{\mathcal{S}}^\ell)+S(\rho_{\mathcal{S}}^m)-S(\rho_{\mathcal{S}}^{\ell-m})\qquad &\text{for}\; m<\ell,\\
2S(\rho_{\mathcal{S}}^\ell)\qquad &\text{for} \; m>\ell.
\label{MI1}
\end{cases}
\eea

\begin{figure}[!ht]
  \begin{center}
  \centering{\bf (a)}\\
   \includegraphics[width=0.8\columnwidth]{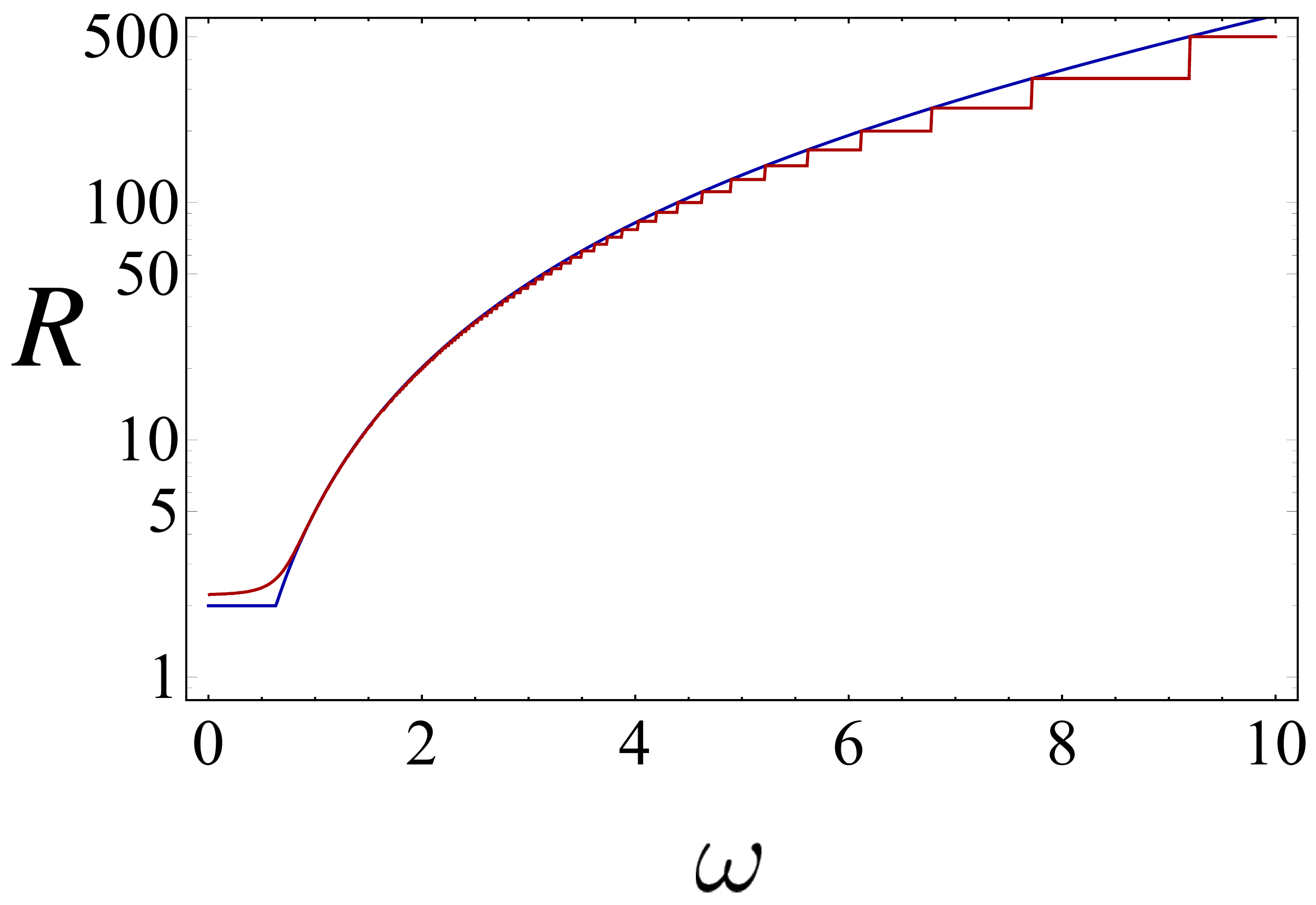}
  \centering{\bf (b)}\\
      \includegraphics[width=0.9\columnwidth]{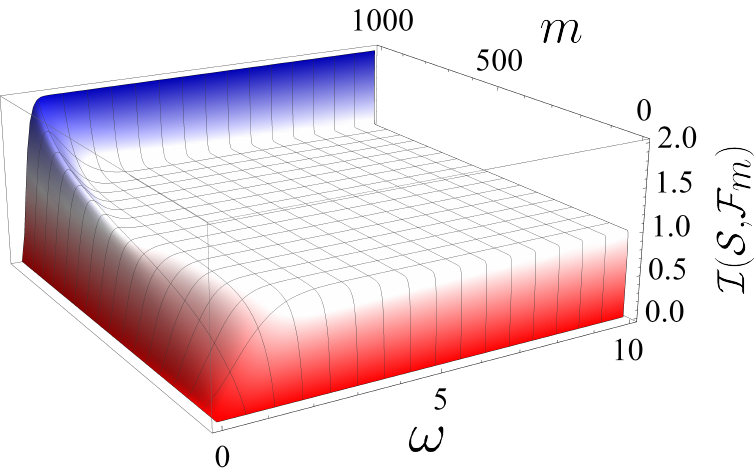}
    \caption{{\bf (a)} The red jagged line shows the redundancy $R$ versus $\om$ after $n=10^3$ collisions and for $\tau=0.05$. The blue curve shows the behavior of the estimated scaling law ${-}n\ln\kappa$. {\bf (b)} We plot the mutual Information $\mathcal{I}(\rho_\mathcal{S},\mathcal{F}_m)$ versus $\om$ and the size of the environmental fraction $m$ for the same parameters as in panel {\bf (a)}. }
    \label{fig_1}
  \end{center}
\end{figure}

In Fig. \ref{fig_1}  the redundancy  $R$ is shown as a function of $\omega$ and its value is compared with  $\Gamma (\tau ) = - n\ln\kappa (\tau )$. The agreement between the two quantities is excellent, confirming our estimate of the redundancy in term of the decay rate $\Gamma$. Note that $R$ is a monotonically growing function of $\omega$ as long as $2\omega\tau \leq \pi /2$.

\section{Dynamical control of the temporal unfolding of \QD} 
\label{controlling}

The study illustrated in the previous Section shows that a collisional model is able to faithfully capture the features linked to the mechanism of temporal unfolding of \QD, and thus account for the building up of redundant information encoding in the state of a multi-element environment. 

We shall now take advantage of such understanding and to show that, by the means of carefully designed (anti-)Zeno-like mechanisms, it is possible to control the spreading of informational redundancy, thus harnessing the unfolding discussed above. Remarkably, the control extends from  dynamical inhibition all the way to the enhancement of the redundancy spreading, which we achieve by engineering the effective dephasing rate of the system. This demonstrates the significant control that can be operated on the fundamental features of \QD, and paves the way to its use for the harnessing of the quantum-to-classical transition. 

\subsection{Inhibiting the spreading of redundancy via the the Quantum Zeno Effect}
\label{Zeno}

We start by addressing the controlled inhibition of \QD. From the discussion in Sec.~\ref{unfolding} it is clear that the workhorse for the achievement of this goal is the control of the information that each ancilla acquires about the system. In turn, this implies the engineering of rate $\Gamma$. A  way of doing this is by engineering a Zeno-like dynamics for the ancillas, which effectively freezes their dynamics during the collisional process. 

The quantum Zeno effect, first pointed out by Misra and Sudarshan in their seminal work~\cite{misra_JoMP1977}, simply states that the continuous observation of a system freezes its evolution. The original argument goes along the following lines: let $|\psi\rangle$ be the initial state of a system evolving according to the Hamiltonian $\hbar {\hat V}$. After a time $t$, the survival probability of the system in its initial state is $|\langle \psi |\exp\{-i{\hat V}t\}|\psi\rangle|^2 \simeq 1 -\langle (\Delta^2\hat V)\rangle t^2 + {\cal O}(t^4)$ with $\Delta^2\hat V=\langle\hat V^2\rangle-\langle\hat V\rangle^2$. Let us assume during time $t$ the system is measured $p$ times: in this case the survival probability becomes $[1 -\langle \Delta^2 {\hat V}\rangle (t/p)^2 ]^p \approx 1$ when $p\rightarrow\infty$. While we do not aim at discussing the considerable body of literature on the Zeno effect produced in the last three decades and spurred by Ref.~\cite{misra_JoMP1977}, here we focus on an early model of constant monitoring put forward by Asher Peres~\cite{peres_AJoP1980} to freeze the ancilla dynamics. 

Let us assume that the ancilllas are no longer two-level systems and instead enlarge the Hilbert space of each $\mathcal{R}_n$ by introducing a third state $\ket{c}$ coupled only to $\ket{a}$ as shown in Fig.\ref{fig_0} {\bf (b)}. This allows us to introduce a dynamics of the subenvironment that is {independent} of the collisional mechanism (we dub it {\it free}), and which we use as our control tool. We thus model each dynamical step through the Hamiltonian 
\bea
\hat H_n=\om \;\hat \sigma_\mathcal{S}^z \otimes e^{i \hat h_n\tau}\left(\ket{a}\bra{b}_{\mathcal{R}_n}+\ket{b}\bra{a}_{\mathcal{R}_n}\right)e^{-i \hat h_n\tau}\label{zenostep}
\eea
with $\hat h_n{=}(\Om/\tau)(\ket{a}\bra{c}_{\mathcal{R}_n}+\ket{c}\bra{a}_{\mathcal{R}_n})$ the free Hamiltonian for the $\ket{a}\leftrightarrow\ket{c}$ transitions and $\Omega$ a dimensionless Rabi rate (we remind that the evolution time $\tau$ due to collisions is {\it set} in our framework, and thus a constant). Eq.~\eqref{zenostep} is thus the interaction Hamiltonian between system and subenvironments written in a reference frame set by the free Hamiltonian $\hat h_n$. In what follows we will require $\Omega/({\omega\tau})\gg1$ so as to freeze the dynamics of the subenvironments.  

\begin{figure}[t!]
  \begin{center}
  \centering{\bf (a)}\\
    \includegraphics[width=0.8\columnwidth]{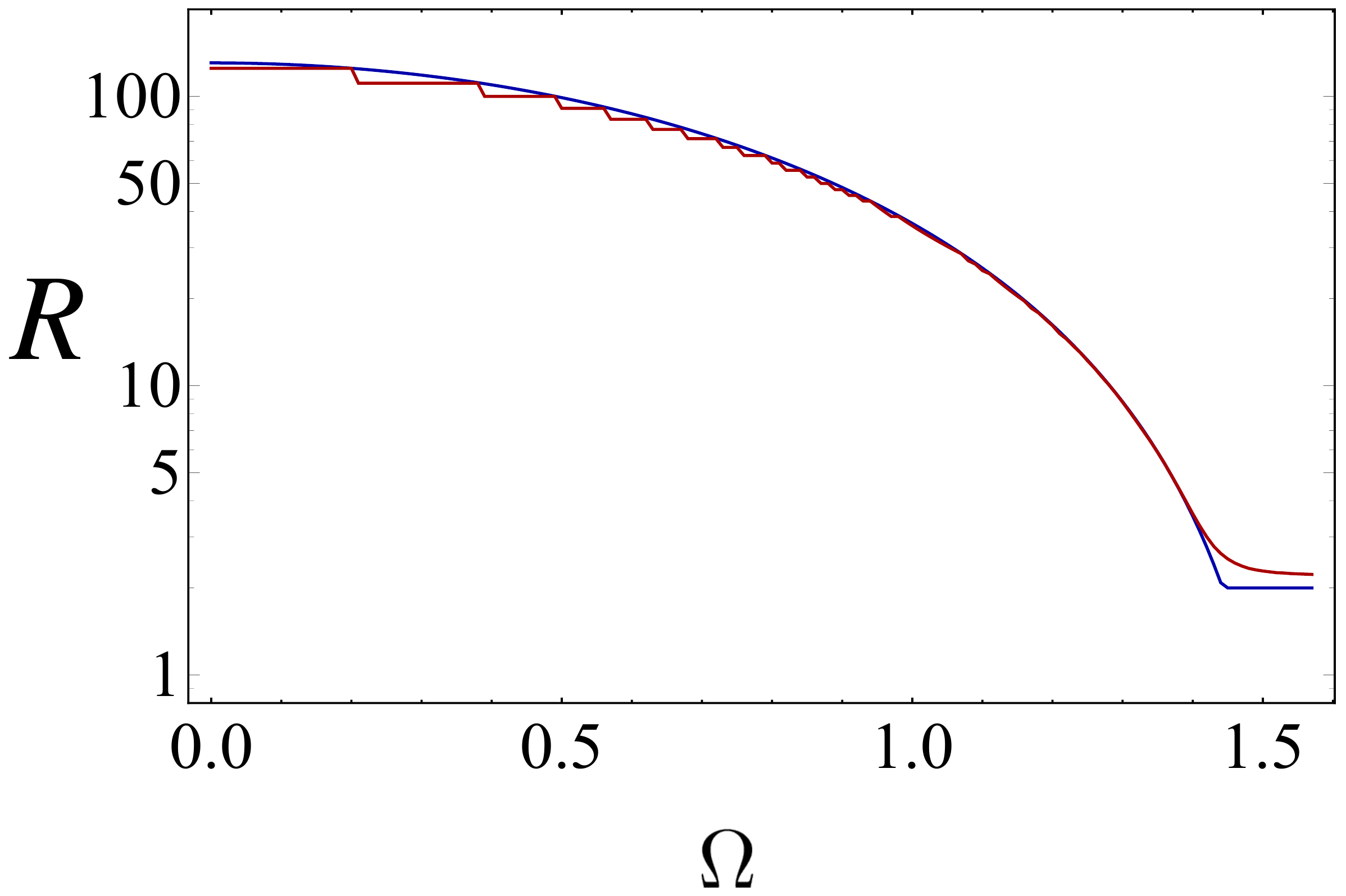}
  \centering{\bf (b)}\\
        \includegraphics[width=0.9\columnwidth]{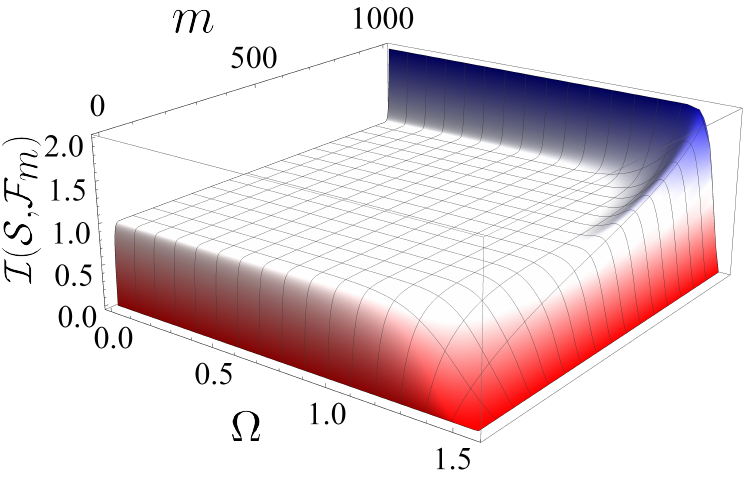}
    \caption{{\bf (a)} We plot the redundancy $R$ (red jagged line) against the dimensionless quantity $\Omega\in[0,\pi/2]$ for $n=10^3$ collisions, each lasting a time $\tau=0.05$,  and for $\om=5$. For $\Omega>\omega\tau$, the redundancy is reduced. the blue curve is shows the behavior of the esitmate ${-}n\log(\kappa)$. {\bf (b)} We show the mutual information $\mathcal{I}(\rho_\mathcal{S},\mathcal{F}_m)$ versus $\Om$ and the fraction $m$ of subenvironments considered.}
    \label{fig_2}
  \end{center}
\end{figure}

In the basis of $\hat\sigma^z_{\cal S}$ the corresponding step evolution operator can be written as
\bea
\hat{U}=\begin{pmatrix}
e^{-i\om\tau \hat M} & {\mathbb O}\\ {\mathbb O} & e^{i\om\tau \hat M}
\end{pmatrix},
\eea
where ${\mathbb O}$ is the null matrix and, in the basis $\{\ket{c},\ket{b},\ket{a}\}_{{\cal R}_n}$, we have
\bea
M=\begin{pmatrix}
 0 & -i \sin\Om & 0 \\
 i \sin\Om & 0 & \cos\Om \\
 0 & \cos\Om & 0 
\end{pmatrix}.\label{Mzeno}
\eea
For $\Om\in[0,\pi/2]$, $M$ has eigenvalues $\lambda_\pm=\pm1$ and $\lambda_0=0$. The associated eigenvectors are
\beq
\begin{aligned}
|{\varphi^\pm}\rangle&=\frac{1}{\sqrt{2}}\left(\cos\Om\ket{a}\pm \ket{b}-i\sin\Om\ket{c}\right),\\
|{\varphi^0}\rangle&=\left(\sin\Om\ket{a}+i\cos\Om\ket{c}\right).
\end{aligned}
\eeq
The reduced density matrix of the system is given again by Eq.~\eqref{reduced}. However, the decoherence factor is modified dynamically as
\bea
\label{zenokappa}
\kappa{=}\sum _{j=\pm,0}\!\! e^{-2 i\om\tau \lambda_j}| \langle{\varphi^j}\vert{a}\rangle|^2{=}\cos(2\omega\tau)\cos^2\Omega{+}\sin^2\Omega,
\eea
which corresponds to a new dephasing rate [cf. Eq.~\eqref{ME1}] reading $\gamma{=}\om^2 \tau\cos^2\Om$. The behavior of the redundancy $R$ associated with such a modified dynamics is shown in Fig.~\ref{fig_2}, which demonstrates that large values of $\Omega$ -- at set values of $\omega\tau$ --  correspond to a depleted information spreading effect: The fact {\it free} Rabi flopping in the $\{\ket{a},\ket{c}\}_{{\cal R}_n}$ subspace induced by a large value of $\Omega$ freeze the evolution of the subenvironments by projecting their state onto $\ket{a}_{{\cal R}_n}$ at a rate that exceeds the effects of any collision. 

\begin{figure}[!b]
	\begin{center}
		\includegraphics[width=\columnwidth]{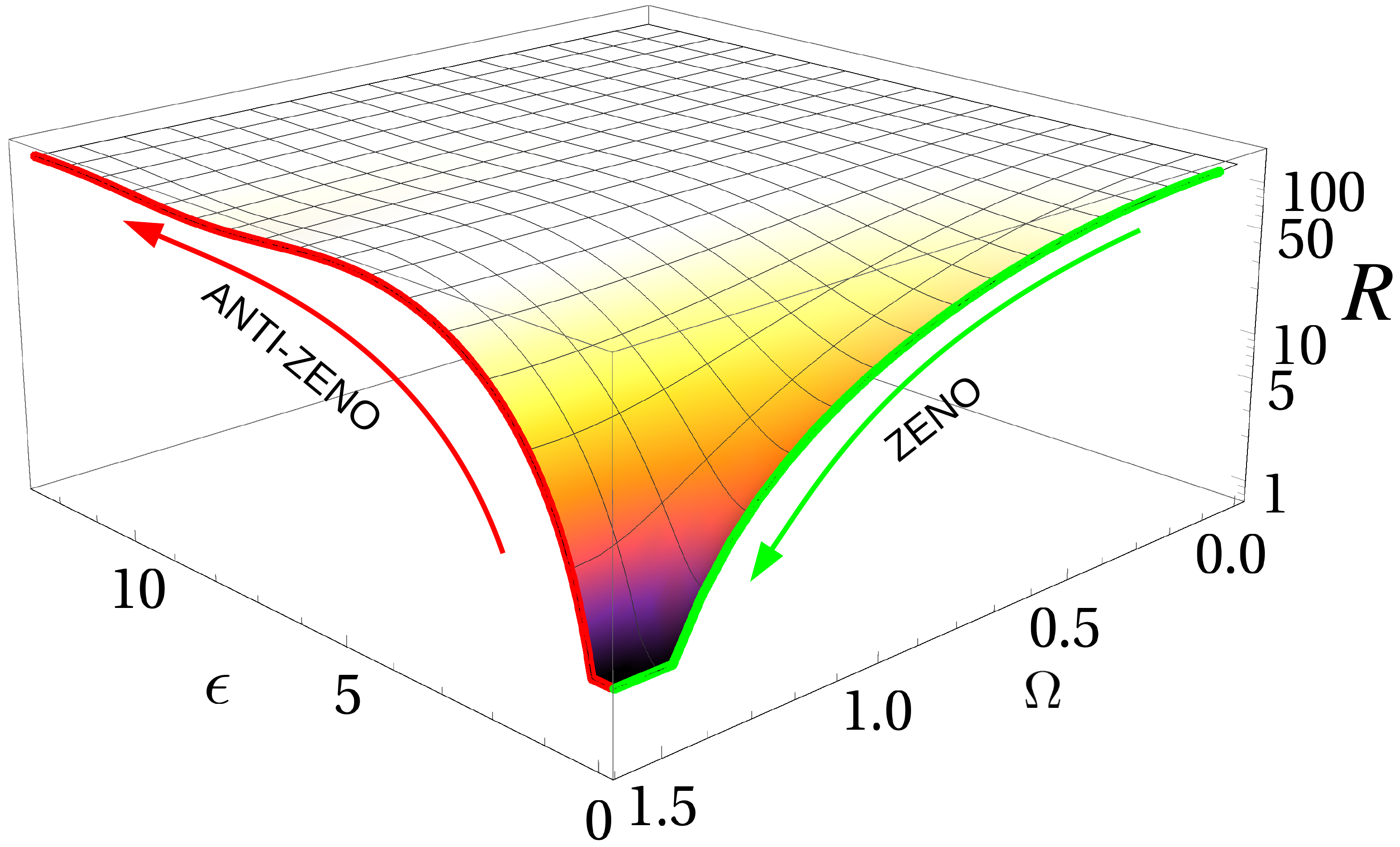}
		\caption{Redundancy $R_\delta$ after $N=10^3$ collisions, $\tau=0.05$ and $\om=5$ versus $\Om$ and $\epsilon$. The green curve is the same as in Fig.~\ref{fig_2}: starting from $\Om\sim 1.5$ and increasing the detuning, the redundancy increase (red curve).
			}
		\label{fig_3}
	\end{center}
\end{figure}

\subsection{Enhancing the spreading of redundancy via the Quantum Anti-Zeno Effect}
\label{AntiZeno}

We now reverse our goal and look for a strategy that allows for the enhancement of the information spreading effect that unfolds \QD. Taking advantage of the analysis presented in Sec.~\ref{Zeno}, we aim at modelling an effective anti-Zeno mechanism that accelerates the rate of collisional spin-flip induced by the ${\cal S}$-${\cal R}_n$ interaction. The solution is remarkably simple. 

We model each step again as in Eq.~\eqref{zenostep}. However, we now consider the possibility to have an energy detuning $2\epsilon$ between $\ket{a}$ and $\ket{c}$, which changes the free Hamiltonian $\hat h_n$ as 
\begin{equation}
h_n{=}-2(\epsilon/\tau)\ketbra{c}{c}_{\mathcal{R}_n}+(\Om/\tau)\left(\ket{a}\bra{c}_{\mathcal{R}_n}+\ket{c}\bra{a}_{\mathcal{R}_n}\right)
\end{equation}
and, in turn, matrix $M$ as
\begin{equation}
M{=}
\begin{pmatrix}
0&\dfrac{i\Om e^{-i\epsilon}\sin \nu}{\nu}&0\\
-\dfrac{i\Om e^{i\epsilon}\sin\nu}{\nu}&0&e^{i\epsilon}\left(\cos\nu{-}\dfrac{i\epsilon\sin\nu}{\nu}\right)\\
0&e^{-i\epsilon}\left(\cos\nu{+}\dfrac{i\epsilon\sin\nu}{\nu}\right)&0
\end{pmatrix},\label{Mantizeno}
\end{equation}
where $\nu=\sqrt{\epsilon^2+\Om^2}$. Following steps akin to those presented in Ref.~\ref{Zeno}, it is straightforward to evaluate the affected dephasing rate and the new redundancy $R$, which now depends on 
\begin{equation}
\kappa=\frac{\Omega ^2 \sin ^2\nu +\cos (2 \tau  \omega ) \left(\nu ^2 \cos ^2\nu+\epsilon
   ^2 \sin ^2\nu\right)}{\nu ^2}.
\end{equation}
This expression goes back to Eq.~\eqref{zenokappa} for $\epsilon\to0$ but leads to a redundancy that can be larger than the one associated with no detuning, as shown in Fig.~\ref{fig_3}, thus demonstrating the possibility to enhance the information spreading.

\section{Conclusions}
\label{conc}

By making use of a collisional approach to open system dynamics, we have shed light onto the temporal unfolding of the redundant encoding of information that is at the basis of the emergence of \QD. Remarkably, this has opened the way to the design of quantum Zeno-based techniques for the control of the information spreading rate, which can be enhanced or depleted by engineering system-environment couplings and suitably arranging the spectrum of the subenvironments. We believe this work draws a pathway to the design of experimental settings able to mimic and thus validate the features and controlled dynamics discussed here. In turn, this will contribute further to the clarification of the role that \QD~plays in the understanding of the quantum to classical transition~\cite{Quanta}. 

\acknowledgments
We are grateful to W. H. Zurek for discussions. We  acknowledge support under PRIN project 2017SRNBRK QUSHIP funded by MIUR,  the EU Collaborative project TEQ (grant agreement 766900), the DfE-SFI Investigator Programme (grant 15/IA/2864), COST Action CA15220, the Royal Society Wolfson Research Fellowship ((RSWF\textbackslash R3\textbackslash183013), the Leverhulme Trust Research Project Grant (grant nr.~RGP-2018-266). G.M.P would like to thank the Kavli Institute for  Theoretical Physics, UC Santa Barbara, for its hospitality supported in part by the National Science Foundation under Grant No. NSF PHY-1748958.

\bibliography{Darw}

\end{document}